\documentclass[sigplan,10pt,screen]{acmart}
\renewcommand\footnotetextcopyrightpermission[1]{}

\usepackage[english]{babel}
\usepackage[utf8]{inputenc}
\usepackage{pmboxdraw}
\usepackage{CJKutf8}
\usepackage{wasysym}
\usepackage[normalem]{ulem}

\usepackage{amsfonts,amsmath,graphicx,amsthm}
\usepackage{dsfont}
\usepackage{cuted}  
\usepackage{xcolor}
\usepackage{setspace}
\usepackage{hyperref}
\usepackage{algpseudocode}
\usepackage{siunitx}
\usepackage{tikz}

\usepackage{pdfpages}
\usepackage{xcolor}
\usepackage{colortbl}

\usepackage{soul}
\usepackage{float}
\usepackage{enumitem}
\usepackage{xspace}
\usepackage{subcaption}
\usepackage{graphicx}
\usepackage{listings}
\usepackage[normalem]{ulem}
\usepackage{fontawesome}

\usepackage[ruled,linesnumbered]{algorithm2e}

\usepackage{stackrel}

\newcommand{\remove}[1]{}

\newcounter{globallinecounter}
\definecolor{mGreen}{rgb}{0,0.6,0}
\definecolor{mGray}{rgb}{0.5,0.5,0.5}
\definecolor{mPurple}{rgb}{0.58,0,0.82}
\definecolor{backgroundColour}{rgb}{0.95,0.95,0.92}

\lstdefinestyle{CStyle}{
    commentstyle=\color{mGreen},
    keywordstyle=\color{magenta},
    numberstyle=\tiny\color{mGray},
    stringstyle=\color{mPurple},
    basicstyle=\footnotesize,
    breakatwhitespace=false,         
    breaklines=true,                 
    captionpos=b,                    
    keepspaces=true,                 
    showspaces=false,                
    showstringspaces=false,
    showtabs=false,                  
    tabsize=2,
    language=C
}

\definecolor{lightgray}{gray}{0.6}
\definecolor{lightblue}{rgb}{0.9,0.9,1}
\definecolor{aqua}{rgb}{0.0, 1.0, 1.0}

\newcommand{\name}{\textsc{Lego}\xspace}

\makeatletter
\newcommand{\nosectionlabel}[2]{%
   \protected@write \@auxout {}{\string \newlabel {#1}{{\textbf{#2}}{\thepage}{#2}{#1}{}} }%
   \hypertarget{#1}{\noindent\textbf{#2.}}
}
\makeatother

\newcommand{\decoder}{decoding system\xspace}  
\newcommand{\Decoder}{Decoding System\xspace}  
\newcommand{\compiler}{compiler\xspace}  
\newcommand{\phyc}{physical controller\xspace}  
\newcommand{\logicc}{logical controller\xspace}

\makeatletter
\newcommand{\definetrim}[2]{%
  \define@key{Gin}{#1}[]{\setkeys{Gin}{trim=#2,clip}}%
}
\makeatother
\definetrim{evaluationtrim}{0 1.1cm 0 0}

\algnewcommand{\logicand}{ {\normalfont \textbf{and}}\xspace}
\algnewcommand{\logicor}{ {\normalfont \textbf{or}}\xspace}
\algnewcommand{\logicnot}{ {\normalfont \textbf{not}}\xspace}

\usepackage{multirow}

\title{\huge \name: QEC \Decoder Architecture for Dynamic Circuits}
\author{\normalsize Yue Wu, Namitha Liyanage and Lin Zhong\\ Department of Computer Science, Yale University, New Haven, CT}

\begin{document}

\settopmatter{printacmref=false}
\settopmatter{printfolios=true}
\maketitle
\pagestyle{plain}

\section{Introduction}

Quantum error correction (QEC) is a critical component of FTQC; the QEC \decoder is an important part of Classical Computing for Quantum or C4Q.
Recent years have seen fast development in real-time QEC decoders~\cite{ueno2021qecool,ueno2022qulatis,das2022lilliput,overwater2022neural,liyanage2023scalable,wu2023fusion,higgott2023sparse,barber2023realtime,vittal2023astrea,alavisamani2024promatch,liyanage2024fpga}.
Existing efforts to build real-time decoders have yet to achieve a critical milestone: decoding dynamic logical circuits with error-corrected readout and feed forward.
Achieving this requires significant engineering effort to adapt and reconfigure the decoders during runtime, depending on the branching of the logical circuit.

We present a QEC \decoder architecture called \name, with the ambitious goal of supporting dynamic logical operations. 
\name employs a novel abstraction called the \emph{decoding block} to describe the decoding problem of a dynamic logical circuit.
Moreover, decoding blocks can be combined with three other ideas to improve the efficiency, accuracy and latency of the \decoder.
First, they provide data and task parallelisms when combined with  \emph{fusion-based decoding}~\cite{wu2023fusion}.
Second, they can exploit the pipeline parallelism inside multi-stage decoders.
Finally, they serve as basic units of work for computational resource management. 

Using decoding blocks, \name can be easily reconfigured to support all QEC settings and to easily accommodate innovations in three interdependent fields: code~\cite{dennis2002topological,bravyi2024high}, logical operations~\cite{kitaev2003fault,litinski2019magic,zhou2024algorithmic,gidney2024magic} and qubit hardware~\cite{bluvstein2024logical,acharya2024quantum,reichardt2024demonstration}.
In contrast, existing decoders are highly specialized to a specific QEC setting, which leads to redundant research and engineering efforts, slows down innovation, and further fragments the nascent quantum computing industry.

\section{\name \Decoder Architecture}\label{sec:architecture}

\begin{figure}[tb]
    \centering
    \includegraphics[width=\linewidth,page=1]{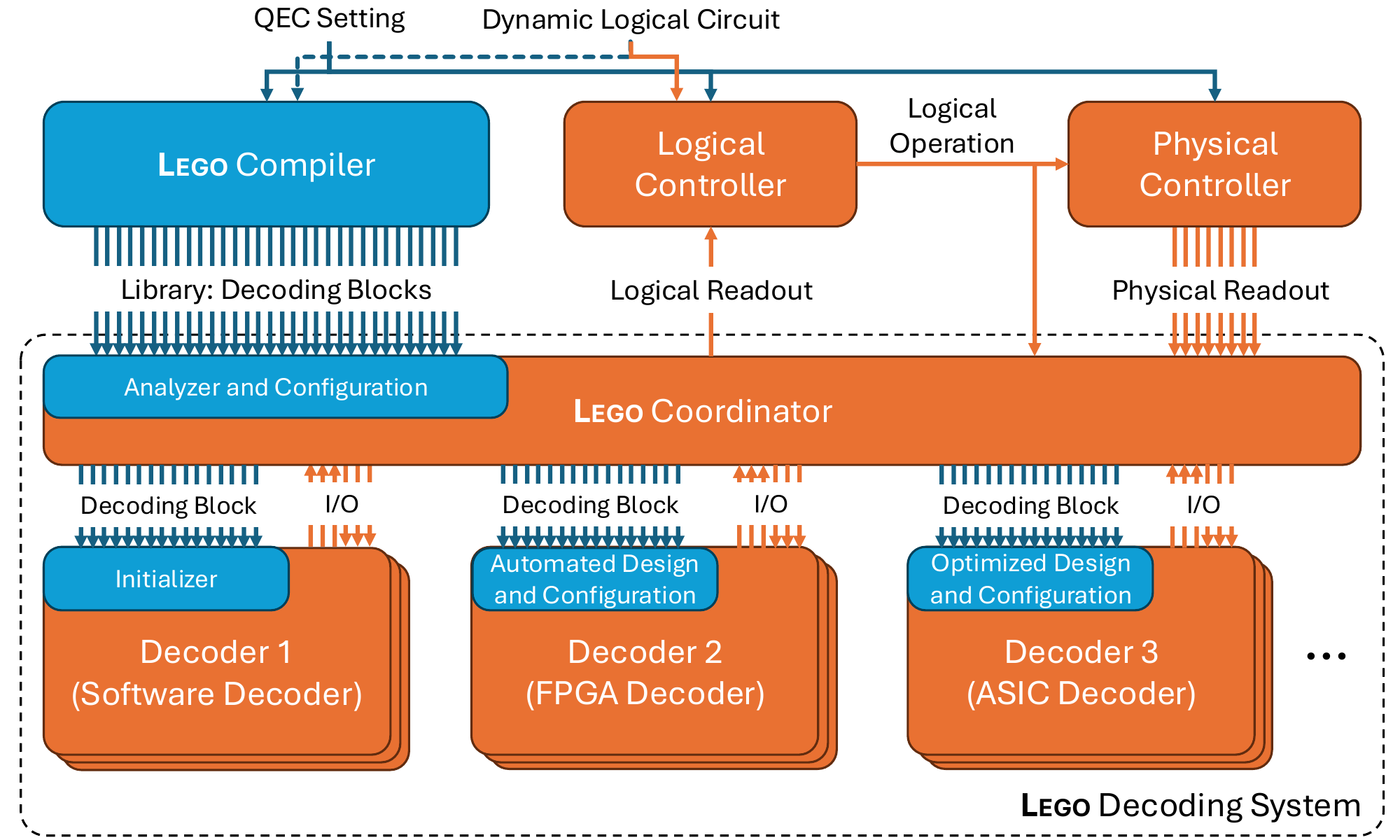}
    \caption{\name \decoder and its interaction with other classical components of an FTQC. The orange and blue blocks represent online and offline components, respectively.}
    \vspace{-2ex}
    \label{fig:architecture}
    \end{figure}

We place the proposed \name \decoder inside the classical computing part of FTQC as illustrated by \autoref{fig:architecture}, with well-defined, technology-agnostic interfaces. 
The input to the quantum computer is a logical circuit specified using a programming language such as OpenQASM. Like a classical software program, the logical circuit consists of operations and conditionals. Due to the use of conditionals, the exact sequence of operations is only known at run-time. \textit{The \phyc} directly talks to quantum hardware, generating control signals and receiving measurements, i.e., physical readouts, and sending them to the \decoder to compute the logical readout. \textit{The \logicc} follows the logical circuit: it receives the logical readout from the \decoder and informs  the \phyc and \decoder what logical operation is the next so that the latter can perform the operation and its QEC decoding, respectively. The \textit{\compiler} takes both the QEC setting and, optionally, the dynamic logical circuit (user program) as input, and generates decoding blocks that can be \emph{merged} into any possible decoding graph of the logical circuit.

QEC decoding systems commonly take physical readout as input and output logical readout; \name takes two additional inputs: (1) the logical operation being executed by the computer, which is known at runtime; and (2) decoding blocks for all logical operations, which are generated offline. 

\name itself is a specialized computer. Hardware-wise, it includes a collection of decoders of varying degrees of specialization. Software-wise, the coordinator functions as a resource manager or operating system that schedules/maps decoding blocks to decoders, potentially with support from the \compiler. 

\section{Decoding Graph and Decoding Block}\label{sec:decoding-graph}

We introduced the notion of decoding graph in~\cite{wu2022interpretation} in which an edge represents an error source and a vertex a detector~\cite{gidney2021stim}, an XOR of a set of stabilizer measurements. 
Because an error source may impact more than two detector readings, edges can be hyperedges and the graph can be a hypergraph. Our key insight is that a decoding graph can precisely describe the decoding problem of many important classes of qubit codes~\cite{kribs2005unified,hastings2021dynamically}.
Many existing QEC decoders accept a decoding graph~\cite{poulin2008iterative,delfosse2022toward,wu2024hypergraph,delfosse2021almost,wu2023fusion,higgott2023sparse} as input.
For a static logical circuit that does not contain any conditionals, one can statically generate the entire decoding graph for its decoding problem because there is a single execution path. 
For a dynamic logical circuit, there can be many possible execution paths, each with a different decoding problem. As the actual execution path is only known at runtime, its decoding graph can only be constructed at runtime, raising a challenge to real-time decoding. \name solves it with the abstraction of the decoding block.

\textbf{Decoding Blocks}:~~When a quantum computer executes a logical operation, the operation
contributes to sources of errors (edges) and detector readings (vertices).
These vertices and edges form a decoding graph $G_1 = (V_1, E_1)$ that describes the decoding problem contributed by this logical operation.
The next logical operation in the execution may contribute another decoding graph $G_2 = (V_2, E_2)$. 
We call $B = V_1\cap V_2$ the combination boundary between $G_1$ and $G_2$. 
$B$ is non-empty if the two operations operate on the same qubit during the same QEC cycle. 
The \decoder must combine $G_1$ and $G_2$ at this boundary.
For a dynamic logical circuit, the \decoder must combine such decoding graphs from logical operations at runtime, adding decoding latency.

\begin{figure}[t]
\centering
\includegraphics[width=\linewidth]{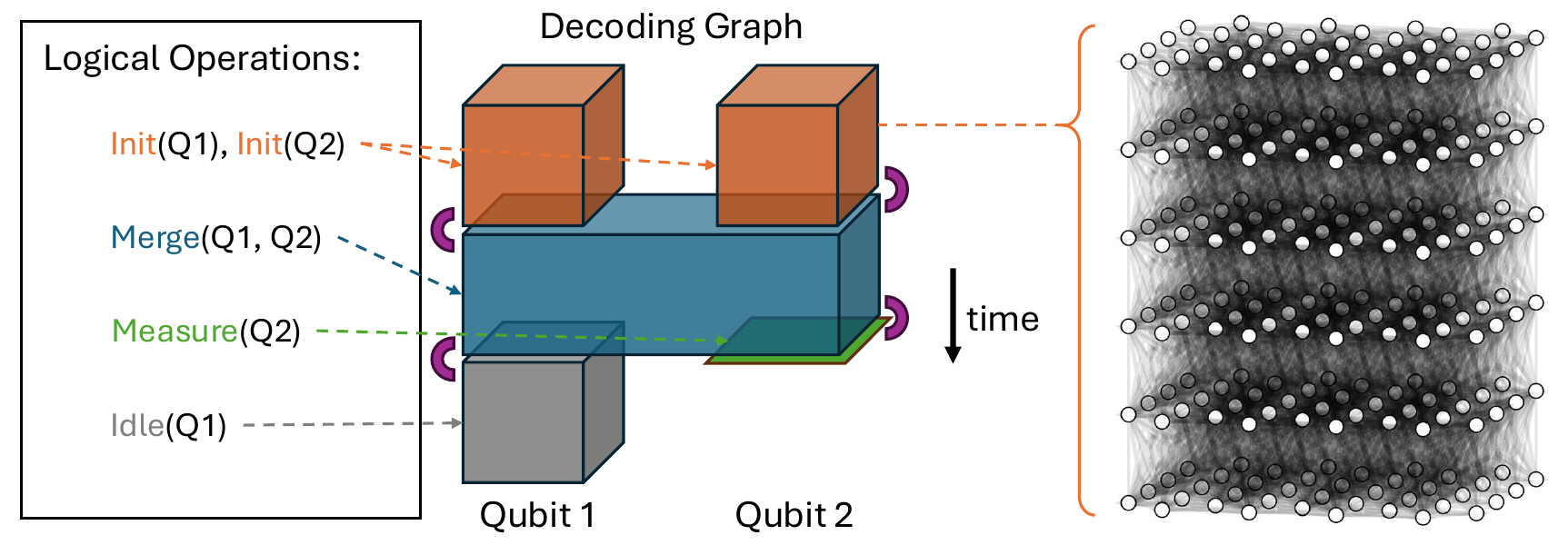}
\caption{The QEC decoding problem of a logical circuit (left) can be described by a decoding graph (middle) as a combination of decoding blocks (right).}
\vspace{-2ex}
\label{fig:example-1}
\end{figure}

We introduce a new abstraction called \emph{decoding block}, or simply \emph{block}, for each logical operation in a dynamic logical circuit. 
Let $O_i, i=1,2...,n$ denote the $i$th operation.
$G_i = (V_i, E_i)$ denote the decoding graph it contributes. Its combination boundary with that of $O_j$ is therefore $B_{ij}=V_i\cap V_j$.
The block for $O_i$ is defined as a tuple: $(G_i, B_i=\{B_{ij} = V_i\cap V_j, i\neq j\})$. That is, it includes $O_i$'s decoding graph and all its combination boundaries with other operations in the same execution. 
An example is shown in \autoref{fig:example-1}.

We say two blocks are of the same type if they  have identical decoding graphs. Logical operations of the same type operating on the same set of qubits will produce blocks of the same type, no matter where they appear in the logical circuit. Blocks of the same type can share the same decoder, providing an opportunity for runtime optimization. 

\textbf{Generating Blocks}:~~Importantly, given a logical circuit, all its blocks can be generated statically, i.e., offline. 
$G_i$ can be generated based on the QEC setting, including the physical layout of logical qubits. To generate $B_i$, a \compiler must enumerate all execution paths and derive the combination boundaries for the operation in each path. 
The \compiler can change the granularity of the blocks, making trade-offs between a large number of small decoding blocks and a small number of large decoding blocks or finding an optimal combination of small and large.
For example, the \compiler can generate a single block for a series of operations in the same execution path. This block will have a large decoding graph but fewer combination boundaries for the \decoder to handle at runtime.
We note that in general, small blocks provide finer granularity for computational parallelism and scheduling flexibility, at the cost of more runtime overhead due to combination.

\textbf{Decoders for Blocks}:~~
In \name, blocks are basic units of work and decoders are basic units of computational resource. A decoder can be completely programmable like a CPU core or FPGA. In this case, it can support any block as long as the necessary program is available. 
A decoder can also be specialized to support a specific decoding graph and therefore, a specific type of blocks.
We can also imagine a more general decoder that can support decoding graphs of certain properties and therefore, support more types of blocks. What types of decoders to develop and include in \name is an important task for the designer.

\section{\Decoder Design}
\label{sec:decoder}

In this section, we elaborate on the potential design opportunities brought by decoding blocks.

\textbf{Fusion-based Decoding}:~~
Decoding blocks conveniently support fusion-based decoding~\cite{wu2023fusion} as their decoding graphs can serve as the partitions with their combination boundaries being the fusion boundaries. 
With fusion-based decoding, each decoding block can be decoded independently, in parallel, and their results are then used to find a solution for the combined decoding graph efficiently without loss of accuracy.
Fusion-based decoding was first supported for the MWPM decoder as a parallelization technique~\cite{wu2023fusion} and was recently generalized to the Union-Find decoder~\cite{liyanage2024multi} and MWPF decoder~\cite{yang2024parallel}.
We hypothesize that other QEC decoders can be adapted for fusion-based decoding. We also note that window decoding~\cite{dennis2002topological} can be used in place of the fusion operation~\cite{wu2023fusion} in fusion-based decoding, albeit less efficiently with redundant computation, less accuracy, and restricted boundary conditions~\cite{iyengar2011windowed,tan2022scalable,skoric2023parallel,bombin2023modular}.

\textbf{Efficient Multi-Stage Decoding}:~~Decoding blocks also support multi-stage decoding efficiently and flexibly.
Multi-stage decoding combines multiple decoders by passing the output of one to the input of another, for better accuracy~\cite{higgott2022fragile,jones2024improved} or reduced bandwidth~\cite{delfosse2020hierarchical}.
However, multi-stage decoders are not suitable for low-latency decoding because a later stage cannot start until all earlier stages finish.
Decoding blocks support pipeline parallelism inside multi-stage decoding with adaptive delayed scheduling.

\textbf{Coordinator}:~~ 
Given a logical operation, the coordinator creates a decoding task, from its decoding block and assigns it to one of the many decoders. 
Because the quantum computer could execute multiple logical operations concurrently and not all decoders can execute all decoding blocks, the coordinator resembles the operating system of a heterogeneous classical computer. On the other hand, QEC decoding brings its own set of challenges due to the tight latency requirement.
Moreover, as the \compiler generates the decoding blocks (and their granularity), it can inform and even collaborate with the coordinator for optimized resource management.

\section{\name Roadmap}\label{sec:roadmap}
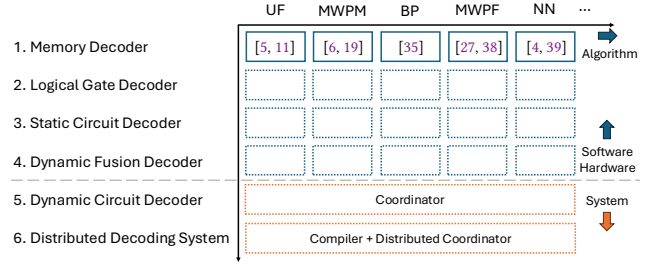
\begin{figure}[t]
    \centering
    \begin{tikzpicture}
    \node[inner sep=0pt] (russell) at (0,0)
        {\includegraphics[width=\linewidth]{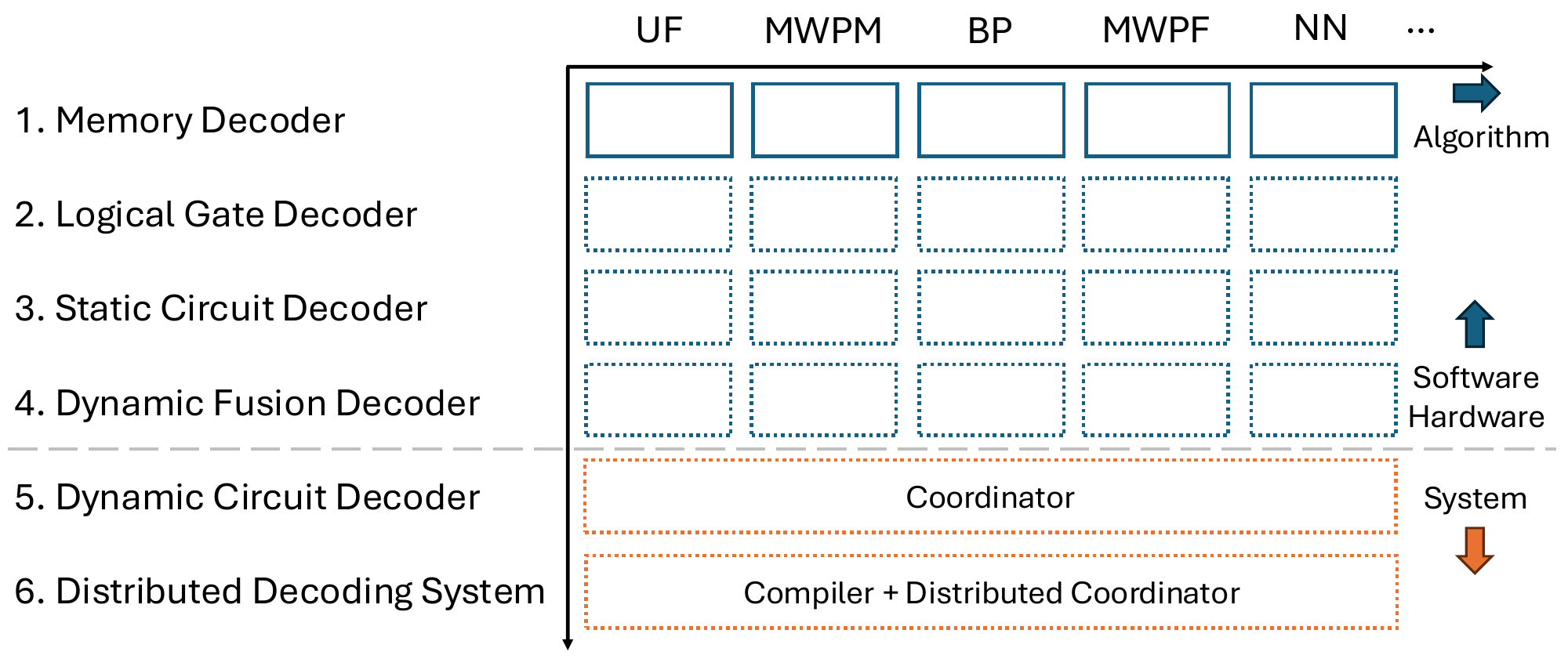}};
    \node[align=center] at (-0.68,1.12) {\tiny\cite{liyanage2023scalable,liyanage2024fpga}};
    \node[align=center] at (0.23,1.12) {\tiny\cite{wu2023fusion,acharya2024quantum}};
    \node[align=center] at (1.13,1.12) {\tiny\cite{higgott2022fragile}};
    \node[align=center] at (2.03,1.12) {\tiny\cite{wu2024hypergraph,hyperion-py}};
    \node[align=center] at (2.93,1.12) {\tiny\cite{overwater2022neural,gicev2023scalable}};
    \end{tikzpicture}
    \caption{\name enables a roadmap of improving decoder capabilities, with three orthogonal research directions.}
    \label{fig:roadmap}
\end{figure}

We present a roadmap of \name, which
defines the levels of capability of a QEC \decoder, as illustrated by 
\autoref{fig:roadmap}.
The first four involve development of individual decoders that doesn't require system-level coordination.
A level-4 decoder supporting fusion-based decoding is considered a valid \emph{fusion-based decoder}.
Starting from level 5, the research focuses the generic decoding system as a whole and is agnostic to concrete choices of decoders.
At level 6, the decoding system becomes scalable for large-scale FTQC.
Once large-scale FTQC is mature, people can then integrate the decoding system with quantum chips.

\textbf{Level 1: Memory Decoder} decodes a small code block, useful for evaluating the basic decoding accuracy and speed.
\textbf{Challenge:} software and hardware optimizations.

\textbf{Level 2: Logical Gate Decoder} decodes individual native logical operations, useful for evaluating the performance beyond memory experiments.
\textbf{Challenge:} adapting decoder to different decoding hypergraphs.

\textbf{Level 3: Static Circuit Decoder} decodes a static logical circuit with an arbitrary number of logical gates.
\textbf{Challenge:} scaling up the decoder and achieving the desired decoding throughput and latency requirements under various circuits.

\textbf{Level 4: Conditional Gate Decoder} dynamically fuses different decoding hypergraphs given an online condition, useful for evaluating real-time logical feed-forward gates.
\textbf{Challenge:} supporting fast fusion-based decoding.

\textbf{Level 5: Dynamic Circuit Decoder} is the first system-level decoder that decodes small-scale dynamic logical circuits on a single machine.
\textbf{Challenge:} building a \name coordinator that routes the physical readouts, manages the decoders, schedules the fusion operations, and outputs logical readouts.

\textbf{Level 6: Distributed Decoding System} leverages decoders from multiple machines to support large-scale logical circuits with more logical qubits and distributed physical readout.
\textbf{Challenge:} building a coordinator that efficiently manages decoder resources in different machines with low-latency communication and synchronization.

The above roadmap indicates that there are three orthogonal but related research directions.
For levels 1 to 4, \emph{algorithm} research designs new decoding algorithms, while \emph{software}/\emph{hardware} research improves the capability of individual decoding algorithms.
From level 5 onward, \emph{system} research focuses on the offline \emph{compiler} and the online \emph{coordinator}.
These research efforts can proceed in parallel, with the aim of accelerating the overall development cycle.

\section*{Acknowledgments}
This work is supported in part by Yale University and NSF MRI Award \#2216030.

\bibliographystyle{lowerCaseTitle.bst}
\bibliography{main}

\begin{thebibliography}{10}

\bibitem{ueno2021qecool}
Yosuke Ueno, Masaaki Kondo, Masamitsu Tanaka, Yasunari Suzuki, and Yutaka
  Tabuchi.
\newblock {QECOOL}: On-line quantum error correction with a superconducting
  decoder for surface code.
\newblock In {\em 2021 58th ACM/IEEE Design Automation Conference (DAC)}. IEEE,
  2021.

\bibitem{ueno2022qulatis}
Yosuke Ueno, Masaaki Kondo, Masamitsu Tanaka, Yasunari Suzuki, and Yutaka
  Tabuchi.
\newblock {QULATIS}: A quantum error correction methodology toward lattice
  surgery.
\newblock In {\em 2022 IEEE International Symposium on High-Performance
  Computer Architecture (HPCA)}. IEEE, 2022.

\bibitem{das2022lilliput}
Poulami Das, Aditya Locharla, and Cody Jones.
\newblock {LILLIPUT}: a lightweight low-latency lookup-table decoder for
  near-term quantum error correction.
\newblock In {\em Proceedings of the 27th ACM International Conference on
  Architectural Support for Programming Languages and Operating Systems}, 2022.

\bibitem{overwater2022neural}
Ramon~WJ Overwater, Masoud Babaie, and Fabio Sebastiano.
\newblock Neural-network decoders for quantum error correction using surface
  codes: A space exploration of the hardware cost-performance tradeoffs.
\newblock {\em IEEE Transactions on Quantum Engineering}, 3:1--19, 2022.

\bibitem{liyanage2023scalable}
Namitha Liyanage, Yue Wu, Alexander Deters, and Lin Zhong.
\newblock Scalable quantum error correction for surface codes using fpga.
\newblock In {\em 2023 IEEE International Conference on Quantum Computing and
  Engineering (QCE)}. IEEE, 2023.

\bibitem{wu2023fusion}
Yue Wu and Lin Zhong.
\newblock {Fusion Blossom}: Fast {MWPM} decoders for {QEC}.
\newblock In {\em 2023 IEEE International Conference on Quantum Computing and
  Engineering (QCE)}. IEEE, 2023.

\bibitem{higgott2023sparse}
Oscar Higgott and Craig Gidney.
\newblock {Sparse Blossom}: correcting a million errors per core second with
  minimum-weight matching.
\newblock {\em arXiv preprint arXiv:2303.15933}, 2023.

\bibitem{barber2023realtime}
Ben Barber, Kenton~M. Barnes, Tomasz Bialas, Okan Bu{\u g}daycı, Earl~T.
  Campbell, Neil~I. Gillespie, Kauser Johar, Ram Rajan, Adam~W. Richardson,
  Luka Skoric, Canberk Topal, Mark~L. Turner, and Abbas~B. Ziad.
\newblock A real-time, scalable, fast and highly resource efficient decoder for
  a quantum computer, 2023.

\bibitem{vittal2023astrea}
Suhas Vittal, Poulami Das, and Moinuddin Qureshi.
\newblock Astrea: Accurate quantum error-decoding via practical minimum-weight
  perfect-matching.
\newblock In {\em Proceedings of the 50th Annual International Symposium on
  Computer Architecture}, ISCA '23, New York, NY, USA, 2023. Association for
  Computing Machinery.

\bibitem{alavisamani2024promatch}
Narges Alavisamani, Suhas Vittal, Ramin Ayanzadeh, Poulami Das, and Moinuddin
  Qureshi.
\newblock Promatch: Extending the reach of real-time quantum error correction
  with adaptive predecoding.
\newblock {\em arXiv preprint arXiv:2404.03136}, 2024.

\bibitem{liyanage2024fpga}
Namitha Liyanage, Yue Wu, Siona Tagare, and Lin Zhong.
\newblock Fpga-based distributed union-find decoder for surface codes.
\newblock 2024.

\bibitem{dennis2002topological}
Eric Dennis, Alexei Kitaev, Andrew Landahl, and John Preskill.
\newblock Topological quantum memory.
\newblock {\em Journal of Mathematical Physics}, 43(9):4452--4505, 2002.

\bibitem{bravyi2024high}
Sergey Bravyi, Andrew~W Cross, Jay~M Gambetta, Dmitri Maslov, Patrick Rall, and
  Theodore~J Yoder.
\newblock High-threshold and low-overhead fault-tolerant quantum memory.
\newblock {\em Nature}, 627(8005):778--782, 2024.

\bibitem{kitaev2003fault}
A~Yu Kitaev.
\newblock Fault-tolerant quantum computation by anyons.
\newblock {\em Annals of Physics}, 303(1):2--30, 2003.

\bibitem{litinski2019magic}
Daniel Litinski.
\newblock Magic state distillation: Not as costly as you think.
\newblock {\em Quantum}, 3:205, 2019.

\bibitem{zhou2024algorithmic}
Hengyun Zhou, Chen Zhao, Madelyn Cain, Dolev Bluvstein, Casey Duckering,
  Hong-Ye Hu, Sheng-Tao Wang, Aleksander Kubica, and Mikhail~D Lukin.
\newblock Algorithmic fault tolerance for fast quantum computing.
\newblock {\em arXiv preprint arXiv:2406.17653}, 2024.

\bibitem{gidney2024magic}
Craig Gidney, Noah Shutty, and Cody Jones.
\newblock Magic state cultivation: growing t states as cheap as cnot gates.
\newblock {\em arXiv preprint arXiv:2409.17595}, 2024.

\bibitem{bluvstein2024logical}
Dolev Bluvstein, Simon~J Evered, Alexandra~A Geim, Sophie~H Li, Hengyun Zhou,
  Tom Manovitz, Sepehr Ebadi, Madelyn Cain, Marcin Kalinowski, Dominik
  Hangleiter, et~al.
\newblock Logical quantum processor based on reconfigurable atom arrays.
\newblock {\em Nature}, 626(7997):58--65, 2024.

\bibitem{acharya2024quantum}
Rajeev Acharya, Laleh Aghababaie-Beni, Igor Aleiner, Trond~I Andersen, Markus
  Ansmann, Frank Arute, Kunal Arya, Abraham Asfaw, Nikita Astrakhantsev, Juan
  Atalaya, et~al.
\newblock Quantum error correction below the surface code threshold.
\newblock {\em arXiv preprint arXiv:2408.13687}, 2024.

\bibitem{reichardt2024demonstration}
Ben~W Reichardt, David Aasen, Rui Chao, Alex Chernoguzov, Wim van Dam, John~P
  Gaebler, Dan Gresh, Dominic Lucchetti, Michael Mills, Steven~A Moses, et~al.
\newblock Demonstration of quantum computation and error correction with a
  tesseract code.
\newblock {\em arXiv preprint arXiv:2409.04628}, 2024.

\bibitem{wu2022interpretation}
Yue Wu, Namitha Liyanage, and Lin Zhong.
\newblock An interpretation of union-find decoder on weighted graphs.
\newblock {\em arXiv preprint arXiv:2211.03288}, 2022.

\bibitem{gidney2021stim}
Craig Gidney.
\newblock Stim: a fast stabilizer circuit simulator.
\newblock {\em Quantum}, 5:497, 2021.

\bibitem{kribs2005unified}
David Kribs, Raymond Laflamme, and David Poulin.
\newblock Unified and generalized approach to quantum error correction.
\newblock {\em Physical review letters}, 94(18):180501, 2005.

\bibitem{hastings2021dynamically}
Matthew~B Hastings and Jeongwan Haah.
\newblock Dynamically generated logical qubits.
\newblock {\em Quantum}, 5:564, 2021.

\bibitem{poulin2008iterative}
David Poulin and Yeojin Chung.
\newblock On the iterative decoding of sparse quantum codes.
\newblock {\em arXiv preprint arXiv:0801.1241}, 2008.

\bibitem{delfosse2022toward}
Nicolas Delfosse, Vivien Londe, and Michael~E Beverland.
\newblock Toward a union-find decoder for quantum {LDPC} codes.
\newblock {\em IEEE Transactions on Information Theory}, 2022.

\bibitem{wu2024hypergraph}
Yue Wu, Lin Zhong, and Shruti Puri.
\newblock Hypergraph minimum-weight parity factor decoder for qec.
\newblock {\em Bulletin of the American Physical Society}, 2024.

\bibitem{delfosse2021almost}
Nicolas Delfosse and Naomi~H Nickerson.
\newblock Almost-linear time decoding algorithm for topological codes.
\newblock {\em Quantum}, 2021.

\bibitem{liyanage2024multi}
Namitha Liyanage, Yue Wu, Emmet Houghton, and Lin Zhong.
\newblock Multi-fpga system for quantum error correction with lattice surgery.
\newblock In {\em 2024 IEEE International Conference on Quantum Computing and
  Engineering (QCE)}. IEEE, 2024.

\bibitem{yang2024parallel}
Liu Yang, Yue Wu, and Lin Zhong.
\newblock Parallel minimum-weight parity factor decoding for quantum error
  correction.
\newblock In {\em 2024 IEEE International Conference on Quantum Computing and
  Engineering (QCE)}. IEEE, 2024.

\bibitem{iyengar2011windowed}
Aravind~R Iyengar, Marco Papaleo, Paul~H Siegel, Jack~Keil Wolf, Alessandro
  Vanelli-Coralli, and Giovanni~E Corazza.
\newblock Windowed decoding of protograph-based ldpc convolutional codes over
  erasure channels.
\newblock {\em IEEE Transactions on Information Theory}, 58(4):2303--2320,
  2011.

\bibitem{tan2022scalable}
Xinyu Tan, Fang Zhang, Rui Chao, Yaoyun Shi, and Jianxin Chen.
\newblock Scalable surface code decoders with parallelization in time.
\newblock {\em PRX Quantum}, 2022.

\bibitem{skoric2023parallel}
Luka Skoric, Dan~E Browne, Kenton~M Barnes, Neil~I Gillespie, and Earl~T
  Campbell.
\newblock Parallel window decoding enables scalable fault tolerant quantum
  computation.
\newblock {\em Nature Communications}, 2023.

\bibitem{bombin2023modular}
H{\'e}ctor Bomb{\'\i}n, Chris Dawson, Ye-Hua Liu, Naomi Nickerson, Fernando
  Pastawski, and Sam Roberts.
\newblock Modular decoding: parallelizable real-time decoding for quantum
  computers.
\newblock {\em arXiv preprint arXiv:2303.04846}, 2023.

\bibitem{higgott2022fragile}
Oscar Higgott, Thomas~C Bohdanowicz, Aleksander Kubica, Steven~T Flammia, and
  Earl~T Campbell.
\newblock Fragile boundaries of tailored surface codes and improved decoding of
  circuit-level noise.
\newblock {\em arXiv preprint arXiv:2203.04948}, 2022.

\bibitem{jones2024improved}
Cody Jones.
\newblock Improved accuracy for decoding surface codes with matching synthesis.
\newblock {\em arXiv preprint arXiv:2408.12135}, 2024.

\bibitem{delfosse2020hierarchical}
Nicolas Delfosse.
\newblock Hierarchical decoding to reduce hardware requirements for quantum
  computing.
\newblock {\em arXiv preprint arXiv:2001.11427}, 2020.

\bibitem{hyperion-py}
Python binding of {Hyperion} library.
\newblock \url{https://pypi.org/project/mwpf}.

\bibitem{gicev2023scalable}
Spiro Gicev, Lloyd~CL Hollenberg, and Muhammad Usman.
\newblock A scalable and fast artificial neural network syndrome decoder for
  surface codes.
\newblock {\em Quantum}, 7:1058, 2023.

\end{thebibliography}


\end{document}